\documentclass[fleqn,12pt,twoside]{article}
\usepackage{espcrc1}

\usepackage{graphicx}

\newcommand{\AmS}{{\protect\the\textfont2
  A\kern-.1667em\lower.5ex\hbox{M}\kern-.125emS}}
\newcommand{\e}{\mbox{e}}

\title{Quantum statistical measurements of an atom laser beam}

\author{M. K. Olsen\address[ACQAO]{ARC Centre of Excellence for Quantum-Atom Optics,\\ 
        School of Physical Sciences, University of Queensland, \\
        Queensland 4072, Australia.}%
        \thanks{Work supported by the Australian Research Council.},
        A. S. Bradley\addressmark ,
        S. A. Haine\address[ANU]{ARC Centre of Excellence for Quantum-Atom Optics, \\
        Department of Physics, Australian National University, \\
        Canberra, Australia.}
        and
        J. J. Hope\addressmark}
       
\begin{document}

\maketitle

\begin{abstract}
We describe a scheme, operating in a manner analogous to a reversed Raman output coupler, for measuring the phase-sensitive quadrature statistics of an atom laser beam. This scheme allows for the transferral of the atomic field statistics to an optical field, for which the quantum statistics may then be measured using 
the well-developed technology of optical homodyne measurement.
\end{abstract}

\section{Introduction}

The development of optical homodyne techniques~\cite{Yuen,mjc} allowed for quadrature measurements of the electromagnetic field, which subsequently led to demonstrations of quantum features such as squeezing~\cite{squeeze}, the Einstein-Podolsky-Rosen (EPR) paradox~\cite{ou} and continuous variable teleportation~\cite{teleport}. As the techniques for the manipulation of Bose-Einstein condensates (BEC) are constantly improving, it may not be long before similar effects will be achieved with bosonic matter waves in the laboratory. Several methods have been proposed for producing highly non-classical BEC states, including correlated atomic pairs and demonstrations of the EPR paradox using molecular dissociation~\cite{KVK}, the transfer of quantum states from an optical field to an atomic field~\cite{jing,fleischhauer,simonjoe}, and the entanglement of an optical field with an atom laser output~\cite{minhacarta}. The successful demonstration of these effects will require some method of performing quadrature measurements on propagating atomic fields, as a matter-wave equivalent of optical homodyne measurements.

In optical homodyne techniques an intense reference field, which may be considered as classical, is mixed on a beamsplitter with the field of interest. Measurement and manipulation of the photocurrents obtained from measurements of the two outputs of the beamsplitter then allow for the quantum statistics of the field to be accessed in a manner not possible with ordinary intensity measurements. This process allows us to measure the optical field quadratures which have a direct analogue in bosonic matter fields. In principle, these could be measured for atomic fields by repeating the steps used in the optical case, but atomic beamsplitters and phase-controlled reference condensates are not simple to produce in the laboratory. We propose a way to circumvent these difficulties by transferring the desired statistics to an optical field before making the measurements.

\section{Atomic quadratures}

The well known field quadratures of quantum optics~\cite{Danbook} are defined as
\begin{equation}
\hat{X}_{a} = \hat{a}+\hat{a}^{\dag},\:\:\mbox{and}\:\: \hat{Y}_{a} = -i\left(\hat{a}-\hat{a}^{\dag}\right),
\label{eq:optquad} 
\end{equation}
where $\hat{a}$ is the photonic annihilation operator. The Heisenberg uncertainty principal then requires that $V(\hat{X})V(\hat{Y})\geq 1$, with an optical field being called "squeezed" when one of these variances is less than one. In an equivalent manner we may use the annihilation and creation field operators for massive bosons to define atomic quadratures,
\begin{equation}
\hat{X}_{\psi}=\hat{\psi}+\hat{\psi}^{\dag}\:\: \mbox{and}\:\: \hat{Y}_{\psi}=-i\left(\hat{\psi}-\hat{\psi}^{\dag}\right),
\label{eq:atomquad}
\end{equation}
which have the same mathematical properties and allow us to define concepts such as squeezing for atomic fields. Although simple to define, it is not obvious how to measure these atomic quadratures. We will show how, under certain conditions, their statistics may be transferred to an optical field and then indirectly measured. Note that all measurements are in fact indirect, as what is registered is always something such as a digital readout of a photocurrent and not the actual quantity being measured. 

\section{Raman incoupling}

\begin{figure}[htb]
\begin{minipage}[t]{75mm}
\includegraphics[width=0.8\columnwidth]{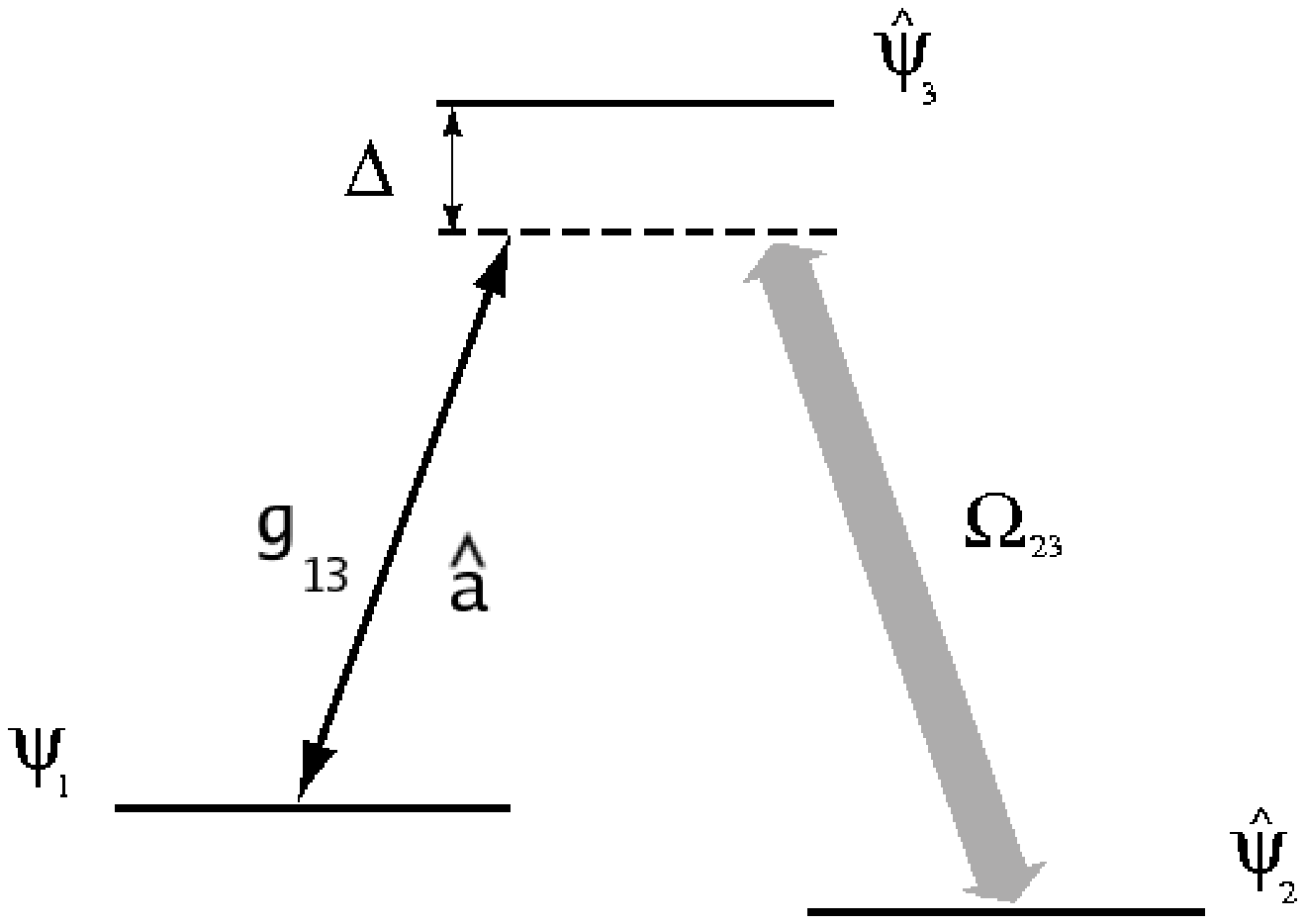}
\caption{Schematic of the Raman incoupling scheme, showing the three atomic levels and the two optical fields.}
\label{fig:levels}
\end{minipage}
\hspace{\fill}
\begin{minipage}[t]{80mm}
\includegraphics[width=0.8\columnwidth]{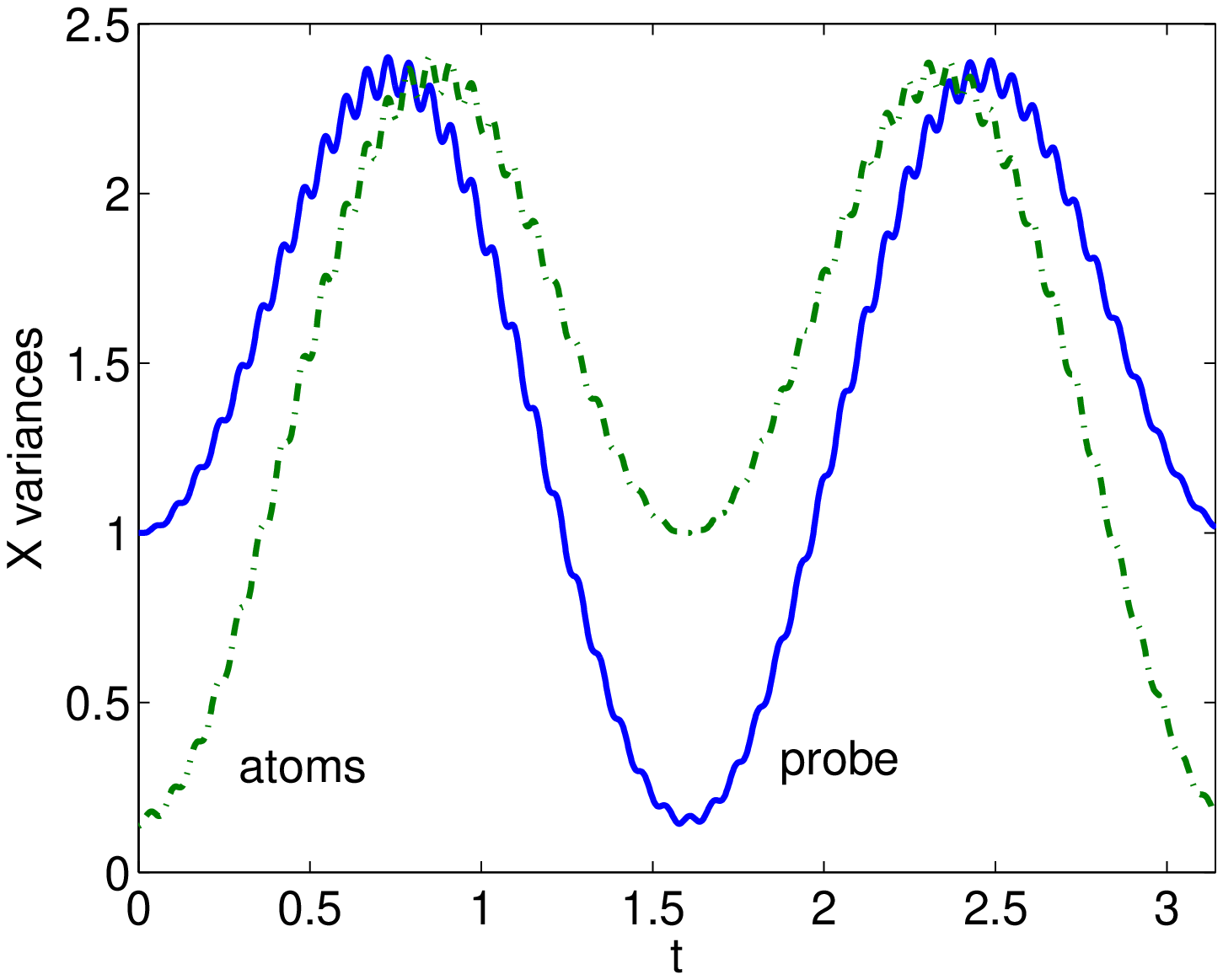}
\caption{Results of stochastic integration of the five-mode model, showing excellent transfer of the atomic statistics to the optical probe.}
\label{fig:Pplus}
\end{minipage}
\end{figure}

Our scheme is an adaptation of the Raman output coupler for atom lasers~\cite{atomlaser}, but with an arrangement of the Raman fields which reverses the coupling, as shown in Fig.~\ref{fig:levels}. At the simplest level, which ignores any spatial effects (for a treatment of these, see Bradley {\em et al.}~\cite{arquivo}), the system may be described by the Hamiltonian
\begin{equation}
{\cal H} = \hbar(\omega_{13}-\Delta)\hat{\psi}_{3}^{\dag}\hat{\psi}_{3}+i\hbar g_{13}\left[\hat{a}\hat{\psi}_{1}\hat{\psi}_{3}^{\dag}-\hat{a}^{\dag}\hat{\psi}_{1}^{\dag}\hat{\psi}_{3}\right] +i\hbar g_{23}\left[\hat{b}\hat{\psi}_{2}\hat{\psi}_{3}^{\dag}-\hat{b}^{\dag}\hat{\psi}_{2}^{\dag}\hat{\psi}_{3}\right],
\label{eq:simples}
\end{equation}
where $\hat{\psi}_{2}$ acts on the atom laser field and $\hat{\psi}_{1}$ is the operator for the highly occupied trapped condensate, while $\hat{\psi}_{3}$ acts on the upper level of the Raman transition. The two annihilation operators $\hat{a}$ and $\hat{b}$ belong to the two laser fields, with $g_{13}$ and $g_{23}$ describing the coupling strength between these and the atomic transitions. These fields are detuned from the upper atomic level by $\Delta$. Note that the atomic levels are not levels of a single atom, but describe different levels of single atoms within interacting atomic fields. 
The next level of approximation, to enable analytical solutions, is to assume that the control field, described by $\hat{b}$, is intense and coherent, so that we may set $g_{23}\langle \hat{b}\rangle=\Omega_{23}$, and that the condensate in level one is large and essentially undepleted, so that we may also set $g_{13}\langle\hat{\psi}_{1}\rangle=\Omega_{13}$. We will set the $\Omega_{ij}$ as real.

\section{Analysis and results}

We may now find the Heisenberg equations of motion for the variables we are treating quantum mechanically as
\begin{equation}
\frac{d\hat{a}}{dt} = -\Omega_{13}\hat{\psi}_{3},\:\:\:\frac{d\hat{\psi}_{2}}{dt} = -\Omega_{23}\hat{\psi}_{3},\:\:\:\frac{d\hat{\psi}_{3}}{dt} = i\tilde{\Delta}\hat{\psi}_{3}+\Omega_{23}\hat{\psi}_{2}+\Omega_{13}\hat{a},
\label{eq:heisenberg}
\end{equation}
where $\tilde{\Delta}=\Delta-\omega_{13}$.
Assuming that $\Delta$ is large, adiabatic elimination of $\hat{\psi}_{3}$ allows us to write the linear Heisenberg equations of motion for $\hat{a}$ and $\hat{\psi}_{2}$,
\begin{equation}
\frac{d\hat{a}}{dt} = -\frac{i\Omega_{13}}{\tilde{\Delta}}\left[\Omega_{23}\hat{\psi}_{2}+\Omega_{13}\hat{a}\right],\:\:\:
\frac{d\hat{\psi}_{2}}{dt} = -\frac{i\Omega_{23}}{\tilde{\Delta}}\left[\Omega_{23}\hat{\psi}_{2}+\Omega_{13}\hat{a}\right],
\label{eq:only2}
\end{equation}
along with the equations for the Hermitian conjugates.
These may be solved analytically to give
\begin{eqnarray}
\hat{a}(t) &=& \frac{\Omega_{23}^{2}+\Omega_{13}^{2}\e^{-i\Omega^{2}t/\tilde{\Delta}}}{\Omega^{2}}\hat{a}(0)+\frac{\Omega_{13}\Omega_{23}\left[\e^{-i\Omega^{2}t/\tilde{\Delta}}-1\right]}
{\Omega^{2}}\hat{\psi}_{2}(0),\nonumber\\
\hat{\psi}_{2}(t) &=& \frac{\Omega_{13}\Omega_{23}\left[\e^{-i\Omega^{2}t/\tilde{\Delta}}-1\right]}{\Omega^{2}}\hat{a}(0)+
\frac{\Omega_{13}^{2}+\Omega_{23}^{2}\e^{-i\Omega^{2}t/\tilde{\Delta}}}{\Omega^{2}}\hat{\psi}_{2}(0),
\label{eq:ansols}
\end{eqnarray}
and their Hermitian conjugates, where we have set $\Omega^{2}=\Omega_{13}^{2}+\Omega_{23}^{2}$. These solutions can be used to show that, at $t=\pi\tilde{\Delta}/\Omega^{2}$,
\begin{equation}
V(\hat{X}_{a})|_{t=\pi\tilde{\Delta}/\Omega^{2}} = \frac{(\Omega_{23}^{2}-\Omega_{13}^{2})^{2}}{\Omega^{4}}V(\hat{X}_{a})|_{t=0}+\frac{4\Omega_{13}^{2}\Omega_{23}^{2}}{\Omega^{4}}V(\hat{X}_{\psi})|_{t=0},
\label{eq:variance}
\end{equation}
where $V(\hat{X}_{\psi})$ is the variance of the $\hat{X}$ quadrature of $\hat{\psi}_{2}$. The same process will work for arbitrary quadrature angle.

Assuming that $V(\hat{X}_{a})|_{t=0}=1$, which holds for either the vacuum or a coherent state, we find
\begin{equation}
V(\hat{X}_{a})|_{t=\pi\tilde{\Delta}/\Omega^{2}} = \frac{(\Omega_{23}^{2}-\Omega_{13}^{2})^{2}}{\Omega^{4}}+\frac{4\Omega_{13}^{2}\Omega_{23}^{2}}{\Omega^{4}}V(\hat{X}_{\psi})|_{t=0},
\label{eq:coerente}
\end{equation}
which involves only known quantities. In the special case of $\Omega_{23}=\Omega_{13}$, we find
\begin{equation}
V(\hat{X}_{a})|_{t=\pi\tilde{\Delta}/\Omega^{2}}=V(\hat{X}_{\psi})|_{t=0},
\label{eq:special}
\end{equation}
in which case there is a complete transfer of the atomic quadrature statistics to the optical field.

We have verified these solutions using stochastic integration of the full five-mode model using the positive-P representation~\cite{P+}, as shown in Fig.~\ref{fig:Pplus}. The initial conditions were $\langle\hat{\psi}_{1}^{\dag}(0)\hat{\psi}_{1}(0)\rangle =  \langle\hat{b}^{\dag}(0)\hat{b}(0)\rangle = 10^{8},\:\langle\hat{\psi}_{3}^{\dag}(0)\hat{\psi}_{3}(0)\rangle=0,\:\langle\hat{\psi}_{2}^{\dag}(0)\hat{\psi}_{2}(0)\rangle = 25014,\:\langle\hat{a}^{\dag}(0)\hat{a}(0)\rangle = 0,\: g_{13}=g_{12}=10^{-3}$ and $\tilde{\Delta}=100$. As can be seen, there is a periodic transfer of statistics between the atomic beam and the optical probe, so that if the interaction time can be tuned to the point of optimal transfer, a complete readout of the atomic variances can be made from the probe field.

\section{Conclusion}

We have shown that it is in principle possible to achieve a perfect transference of quantum statistics from an atomic field to an optical field. A more complicated one-dimensional analysis shows that the scheme works well if certain conditions such as momentum matching can be achieved. Spontaneous emission is not a problem because we operate in a regime where the lasers are detuned from the excited level. Further work is being undertaken to study the behaviour in higher dimensions and also whether there needs to be an established initial phase between the atomic beam and the trapped condensate.

\end{document}